\documentclass[conference]{IEEEtran}
\IEEEoverridecommandlockouts

\usepackage{cite}
\usepackage{amsmath}
\usepackage{algorithmic}
\usepackage{graphicx}
\usepackage{textcomp}
\usepackage{xcolor}
\usepackage{tabularx,booktabs}
\usepackage{multirow}
\usepackage{multicol}
\usepackage{subcaption}
\usepackage{cleveref}
\usepackage{balance}
\usepackage{lastpage}
\usepackage{tabulary}
\usepackage{bbm}
\usepackage{amsmath}
\usepackage{algorithm}  
\usepackage{fancyhdr} 
\newcolumntype{C}{>{\centering\arraybackslash}X} 
\newcolumntype{L}{>{\raggedright\arraybackslash}X}

\def\BibTeX{{\rm B\kern-.05em{\sc i\kern-.025em b}\kern-.08em
    T\kern-.1667em\lower.7ex\hbox{E}\kern-.125emX}}
    
\graphicspath{{images/}}

\begin{document}


\title{Smart Home Energy Management: Sequence-to-Sequence Load Forecasting and Q-Learning\\
}

\author{Mina Razghandi$^\dag$, Hao Zhou$^\ddag$, Melike Erol-Kantarci$^\ddag$, and Damla Turgut$^\dag$\\
{$^\dag$Department of Computer Science, University of Central Florida}\\
{$^\ddag$School of  Electrical Engineering and Computer Science, University of Ottawa }\\
mrazghandi@knights.ucf.edu, turgut@cs.ucf.edu, \{hzhou098, melike.erolkantarci\}@uottawa.ca \\
\vspace{-6 mm}
}

\renewcommand\floatpagefraction{.9}
\renewcommand\topfraction{.9}
\renewcommand\bottomfraction{.9}
\renewcommand\textfraction{.1}
\setcounter{totalnumber}{50}
\setcounter{topnumber}{50}
\setcounter{bottomnumber}{50}
\definecolor{cadmiumgreen}{rgb}{0.0, 0.42, 0.24}
\maketitle

\thispagestyle{fancy} %
      \lhead{} 
      \chead{Accepted by 2021 IEEE Global Communications Conference, \copyright2021 IEEE 
 } 
      \rhead{} 
      \lfoot{} 
      \cfoot{\thepage} 
      \rfoot{} 
      \renewcommand{\headrulewidth}{0pt} 
      \renewcommand{\footrulewidth}{0pt} 
\pagestyle{fancy}

\begin{abstract}
A smart home energy management system (HEMS) can contribute towards reducing the energy costs of customers; however, HEMS suffers from uncertainty in both energy generation and consumption patterns. In this paper, we propose a sequence to sequence (Seq2Seq) learning-based supply and load prediction along with reinforcement learning-based HEMS control. We investigate how the prediction method affects the HEMS operation. First, we use Seq2Seq learning to predict photovoltaic (PV) power and home devices' load. We then apply Q-learning for offline optimization of HEMS based on the prediction results. Finally, we test the online performance of the trained Q-learning scheme with actual PV and load data. The Seq2Seq learning is compared with VARMA, SVR, and LSTM in both prediction and operation levels. The simulation results show that Seq2Seq performs better with a lower prediction error and online operation performance.
\end{abstract}


\section{Introduction}
\label{Introduction}

A smart home energy management system (HEMS) can provide an efficient way to use energy by reducing costs. HEMS controller automatically monitors, controls, and manages home devices, including solar panels, smart appliances, and energy storage systems. For example, the controller may shift some load from the peak energy price period to the off-peak period by controlling the on/off status of smart appliances, e.g., washing machine and dishwasher, and reduce the electricity costs \cite{b2}. However, the HEMS is highly affected by uncertainties. For energy generation, the output power of solar panels changes with time and weather. On the other hand, energy consumption depends on consumers' living habits, which change from one customer to the other \cite{b1}. 

Machine learning and deep learning-based models are becoming more common as a way to solve the limitations of statistical approaches. Deep learning models are known for learning from historical data and making future decisions to handle these non-linear relationships. Consequently, employing deep learning models in aggregated load demand forecasting has received much attention during past years~\cite{li2020designing, li2021short}. Forecasting energy demand for appliances in a household whose use is largely determined by user preferences, such as washing machines, dishwashers, and other appliances, remains a challenge. Deep learning models, such as the Long Short-Term Memory (LSTM), are better suited to predicting appliance energy consumption during the day accurately using smart meter data~\cite{razghandi2020residential}. More sophisticated techniques, such as sequence-to-sequence learning, can also be used to produce more reliable outcomes with fewer contextual data, such as outdoor temperature.

Meanwhile, reinforcement learning (RL) techniques provide opportunities for HEMS control. Given the uncertainty of energy generation and consumption, it is hard for traditional methods to build a dedicated optimization model. However, this complexity is avoided in model-free RL by a uniform Markov decision process (MDP). The agent interacts with the environment and gets rewards by choosing different actions. By exploring possible action combinations, the agent learns the best action sequence to maximize the long-term reward.

Although various prediction methods are proposed in the literature, their comparison is mainly limited with the metrics level, such as Root Mean Squared Error (RMSE), and does not consider the consequence of prediction errors on control systems. 

In this paper, we investigate the performance of control operations in relation to various prediction techniques. First, we compare several forecasting algorithms, including Vector Auto-Regressive Moving Average (VARMA), Support Vector Regression (SVR), Long-Short Term Memory (LSTM), and sequence-to-sequence learning (Seq2Seq). Q-learning is then applied for HEMS, which utilizes the forecasting results for offline training. Finally, we test the trained Q-learning with real-world data and observe how different prediction methods affect the operation cost.

The rest of this paper is organized as follows. Section~\ref{RelatedWork} is the related work. Section~\ref{ProposedModel} presents the system architecture and the proposed method, including the sequence-to-sequence learning forecasting and Q-learning based smart home HEMS. Section~\ref{Experiments} describes the simulation settings and results, and Section~\ref{Conclusion} concludes the paper.

\section{Related Work}
\label{RelatedWork}
Statistical approaches, machine learning-based, deep learning-based, and hybrid models have been mostly used to forecast short-term load in residential buildings. Most common statistical forecasting methods include Autoregressive Integrated Moving Average (ARIMA)~\cite{nepal2020electricity},
Gray Model (GM)~\cite{bian2020research},
and the Kalman Filtering method~\cite{aly2020proposed}.
Most of the works in the literature focus on daily overall load demand on the building or house level. Li et al.~\cite{LI2021116509} propose a Convolutional Long Short-Term Memory based neural network with selected autoregressive features to improve floor-level load prediction accuracy for residential buildings. Haq et al.~\cite{HAQ20201099} classify electrical appliance load into daily, weekly, monthly, and total energy consumption and use a hybrid machine learning method for load forecasting and peak demand. Shomski et al.~\cite{skomski2020sequence} propose a sequence-to-sequence learning method to predict the total energy consumption of commercial buildings with 1 hour and 15 minutes resolution. 

Lu et al.~{\cite{b3}} propose an hour-ahead demand response method for HEMS to minimize the energy cost, which involves a neural network for energy price prediction. An IoT-based self-learning HEMS, proposed by Li et al.~\cite{b4}, utilizes a long-short-term-memory network to predict energy price. A multi-agent RL-based method for HEMS is introduced by Xu et al.~\cite{b5} to reduce energy cost, in which the feed-forward neural network will predict energy price and solar generation. Although various prediction and operation methods are proposed for HEMS, the relationship between load and solar generation prediction along with operation is not investigated, e.g., how different load/supply forecasting methods will affect the operation results. This paper differs from existing works in two aspects. First, an advanced forecasting method, namely sequence-to-sequence learning, is applied for a device-level load prediction using real-world data. Second, based on several baseline algorithms, we further investigate how the prediction accuracy will affect HEMS at the operation level.

\section{System Model and Proposed Prediction and Control Scheme}
\label{ProposedModel}

\subsection{Sequene-to-Sequence Encoder/Decoder Model}
Sustkever et al.~\cite{sutskever2014sequence} presented an end-to-end method for encoding an input sequence to a fixed-length vector and then decoding the vector to produce an output sequence, named sequence-to-sequence learning. They used a multilayered LSTM for both encoder and decoder networks. This network architecture is proposed with the primary goal of managing length differences between phrases while translating from one language to another, which deep neural networks cannot do because they require the dimensionality of the inputs and outputs to be specified and defined.

Let $(x_{1}, x_{2},...,x_{n})$ be the input sequence fed into the encoder, $v$ (fixed-length vector) the
final hidden state of the LSTM encoder network, and $(y_{1}, y_{2},...,y_{m})$ be the corresponding output sequence, where $n$, history window, and $m$, predicting window, may vary, the LSTM decoder network, takes $v$ as the initial hidden state and computes the conditional probability of:
\vspace{-2mm}
\begin{equation}
    p\left ( y_{1},...,y_{m}\vert x_{1},...,x_{n}\right)=\prod_{t=1}^{m}p\left (y_{t}\vert v, y_{1},...,y_{t-1}\right)
    \label{eq:decoderprob}
\end{equation}
\vspace{-2mm}

\subsection{Load Forecasting with sequence-to-sequence Learning}
Inspired by the Sustkever network, we used a sequence-to-sequence learning model for smart home appliances load forecasting. The load forecasting problem can be considered analogous to translating from one language to another in that the history window and the prediction window are not necessarily the same. Also, sequence-to-sequence learning has shown the potential to produce positive outcomes~\cite{razghandi2020Short}. 

\begin{figure*}
    \centering
        \centering
        \includegraphics[width=0.65\textwidth]{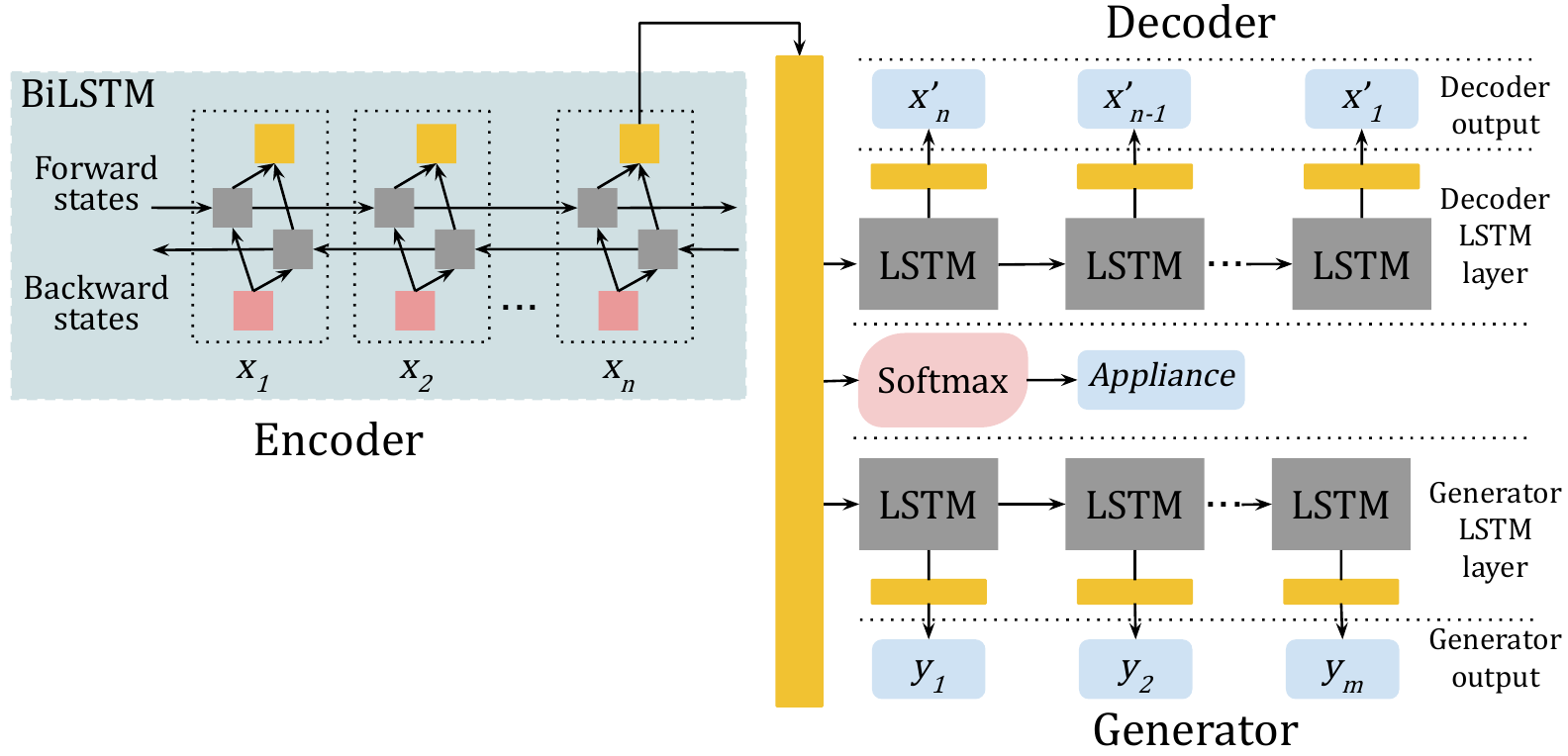}
        \caption{Sequence-to-sequence architecture: encoder (left) with one Bi-LSTM layer, decoder (upper right) and generator (bottom right) with one LSTM layer.}
        \label{fig:model}
        \vspace{-5mm}
\end{figure*}

To perform load forecasting, the proposed model, as depicted in Fig.~\ref{fig:model}, uses the load demand history over the past 24 hours without specifying the type of appliance and predicts load demand for the hour ahead for each appliance. We have used a Bi-directional LSTM (Bi-LSTM) layer in our encoder module. Bi-LSTMs are a type of Bi-directional Recurrent Networks (Bi-RNN)~\cite{schuster1997bidirectional} and benefit from using two independent LSTMs reading input data both in forward and backward directions. In the case of load forecasting, this architecture offers prominent perspectives for the network to understand and preserve underlying relationships from the past and future, for example, on-off time, and encode that information into the fixed-length vector. The decoder module's LSTM layer takes the fixed-length vector and regenerates the input sequence in reverse order, emphasizing recent data. Since the appliance type is not given as an input, the LSTM decoder predicts it, forcing the network to learn each appliance usage pattern. Finally, the LSTM layer in the generator module forecasts the load demand for the prediction window. 

\subsection{Q-learning based HEMS control}
In this section, we introduce a centralized HEMS model that uses the prediction methods of the previous section as inputs. The smart home consists of PV panels, loads, and an energy storage system (ESS). Based on the PV power and load prediction results, the controller aims to maximize the profit of smart home operations. The system is described as following:

\begin{footnotesize}
\vspace{-15pt}
\begin{subequations}\label{e2:main}
\begin{align}
& \text{max} && \sum_{t=1}^{T}(P_{t}^{E}+P_{t}^{PV}-P_{t}^{L})(\alpha p^{sell}_{t}+(1-\alpha)p^{buy}_{t})& \tag{\ref{e2:main}} \\
& \text{s.t.} &&\alpha= \mathbbm{1}\{P_{t}^{E}+P_{T}^{PV}-P_{t}^{L}>0\} & \label{e2:d}  \\
&             && P_{t}^{E}=P_{char}q_{t} & \label{e2:c}\\
&             && S_{t+1}= S_{t}-\frac{P_{t}^{E}}{C^{E}} & \label{e2:f}\\
&             && S_{min}\leq S_{t} \leq S_{max} & \label{e2:g}\\
&             && P_{t}^{L}=\sum_{l=1}^{D}P_{l}G_{l,t} & \label{e2:e}\\
&             && w_{l,t}\leq w_{l,max} & \label{e2:h}
\end{align}
\end{subequations}
\end{footnotesize}
\noindent
where $P_{t}^{E}$, $P_{t}^{PV}$, and $P_{t}^{L}$ denote the power of ESS, PV and load, $p^{sell}_{t}$ and $p^{buy}_{t}$ denote the price of selling and buying electricity, respectively. $\mathbbm{1}$ is an indicator function. $\alpha = 1$ when $P_{t}^{E}+P_{t}^{PV}-P_{t}^{D}>0$ to sell energy, otherwise $\alpha = 0$ to buy energy. $P_{char}$ is the charging power of ESS, $C^{E}$ is ESS capacity, $q_{t} = -1,0,1$ when charge, unchanged and discharge. Constraint (\ref{e2:f}) is the ESS state of charge (SOC) updating constraints, and (\ref{e2:g}) is the SOC upper and lower limit constraint. $P_{l}$ is the power of home device $l$, $D$ is the number of devices, and $G_{l,t}$ is the on/off status. In this work, we consider the demand side management, in which the controller can defer the operation time of some devices without affecting customer comfort. However, the devices have a waiting time limit shown as (\ref{e2:h}), in which $w_{l,t}$ is the waiting time of device $l$, and $w_{l,max}$ is the max waiting time.

Here, we use Q-learning-based control for HEMS operation. The agent state is defined as $\{t,S_{t},w_{l,t}\}$, and the action is $\{q_{t},G_{l,t}\}$. In this algorithm, we first perform prediction-based learning, which utilizes the prediction results of Seq2Seq learning. The trained action combinations are applied in actual PV and load data to verify the performance. The Q-learning-based algorithm is summarized in Algorithm~\ref{Q-LearningAlgo}.

\setlength{\textfloatsep}{5pt}
\begin{algorithm}[!t]
	\caption{Q-learning for Smart Home Economic Dispatch}
	\begin{algorithmic}[1]
		\STATE {\textbf{Prediction-based training:}}
		\STATE \textbf{Initialize:} Q-learning and smart home parameters
		\STATE {Load Seq2Seq PV and load prediction results}
		\FOR{$episode=1$ to $E$}
		\FOR{$t=1$ to $T$}
		\STATE{With probability $\epsilon$ choose action $a$ randomly. Otherwise, $a=arg\;max (Q(s,a))$} 
		\STATE {Agent calculates reward based on equation (\ref{e2:main})}. 
		\STATE {Update agent state $\{t,S_{t},w_{l,t}\}$ and Q-value:} 
		\STATE {$Q(s,a)=(1-\alpha)Q(s,a)+\alpha(r+\gamma max Q(s',a'))$}
		\ENDFOR
		\ENDFOR
		\STATE {Get predicted reward}
		\STATE {\textbf{Test with actual data:}}
		\STATE {Load real PV and load data}
		\STATE {Apply the trained Q-learning and get actual reward}
	\end{algorithmic}
	\label{Q-LearningAlgo}
\end{algorithm}

\section{Evaluation Study}
\label{Experiments}

\subsection{Experiment Setup}
\label{Experimentsetup}
We use a publicly available dataset from iHomeLab RAPT~\cite{huber2020residential}, which contains solar electricity production and domestic electricity consumption measurements for five houses located in Switzerland from April 2016 to August 2019 with a sampling frequency of five minutes. We focus on a residential-detached House D since it includes data on PV production, deferrable loads such as dishwashers, tumble dryers, High Fidelity (HiFi) systems and routers, washing machines, and total power consumption.

As a baseline for testing the performance of the proposed model, we used four load forecasting methods from the literature:

\begin{itemize}
    \item Vector Auto-Regressive Moving Average (VARMA): This model is a combination of Vector Auto-Regressive (VAR) and Vector Moving Average (VMA) models. The VAR model makes predictions assuming that the predicted value is a linear combination of past with random noise. The VMA model assumes the predicted value to be a linear combination of previous errors, expected value, and random noise term.
    \item Support Vector Regression (SVR): SVR is a Support Vector Machine variant used to solve regression problems, where there is a nonlinear relationship between inputs. SVR will map the input data to a higher dimension space in which it is linearly separable using kernel functions and fit a hyperplane with maximum margin to it.
    \item Long Short-Term Memory Neural Network (LSTM): LSTM network is a variant of RNNs that aims to solve RNNs shortcomings for long data sequences. LSTM gating mechanism (input gate, forget gate, output gate) prevents exploding and vanishing gradient issues. Here we have used a two-layer LSTM.
    \item Seq2Seq Learning: This is described in Section~\ref{ProposedModel}.
\end{itemize}

We assume that the capacity of ESS in a smart home is 16 k·Wh, and the charge power is 4kW. The energy trading price is given in~\cite{b6}, where peak/off-peak pricing is applied. The Q-learning discount factor is 0.99, and the learning rate is 0.95. The initial $\epsilon$ value is 0.1, and it decreases with iterations. The prediction-based training contains 7000 iterations, and then it is applied to actual data.

\subsection{Performance Indices}
\label{performanceIndices}
We picked three of the most common error measures in the literature to determine the accuracy of forecasts produced by the proposed Seq2Seq forecasting model and the other baseline models. The Weighted Mean Absolute Percentage Error (wMAPE) is defined as Equation~\ref{eq:wmape} and is a variant of Mean Absolute Percentage Error that overcomes the divide by zero that occurs in MAPE when the base is zero. Another commonly used point error measure that is scale-dependent is Root Mean Square Error (RMSE). We have also used Normalized Root Mean Square Error (nRMSE) to compare the models' performance across different appliances with different scales. RMSE and nRMSE are defined according to Equations ~\ref{eq:rmse}--\ref{eq:nrmse}, where $\mathit{n}$ is the total number of predictions, $\mathit{y_{j}}$ is the actual value and $\mathit{\hat{y_{j}}}$ is the predicted value, $\mathit{max(y_{j})}$ and $\mathit{min(y_{j})}$ refer to the maximum and minimum energy consumption recorded for the appliance $\mathit{j}$ in the data, respectively.
\begin{equation}
  wMAPE = \frac{\sum_{i=1}^{n} \vert y_{j} - \hat{y_{j}}\vert}{\sum_{i=1}^{n} \vert y_{j}\vert}
  \label{eq:wmape}
\end{equation}
\begin{equation}
  RMSE = \sqrt{\frac{1}{n}\sum_{i=1}^{n}(y_{j} - \hat{y_{j}})^{2}}
  \label{eq:rmse}
\end{equation}
\begin{equation}
  nRMSE = \frac{RMSE}{max(y_{j})-min(y_{j})}
  \label{eq:nrmse}
\end{equation}

\subsection{Data Preprocessing}
\label{datapreprocessing}
 
The dataset contains energy consumption data with a  sampling frequency of five minutes. We smooth out the energy consumption of appliances over $T$=10 minutes without compromising generality. The selected house has about 594 days of data after data processing. We divide data into training, validation, and test sets using a 70\%, 20\%, and 10\% ratio. Since each appliance has its energy consumption fluctuation range, spanning from 0W to 15000W, it is critical to normalize data before feeding it to predictor models for the best performance.

Historical energy consumption data of a household is a rich source of knowledge that reveals profound insights into the activities and preferences of its inhabitants, which is valuable to HEMS systems. Hence, we used a sequence of 144 data points (from the previous 24 hours at 10-minute intervals) as input data, along with the day of the week.

\subsection{Prediction Results}
\label{Results}
\begin{figure*}
    \centering
    \vspace{-5mm}
    \begin{subfigure}{0.65\textwidth}
        \includegraphics[width=\linewidth]{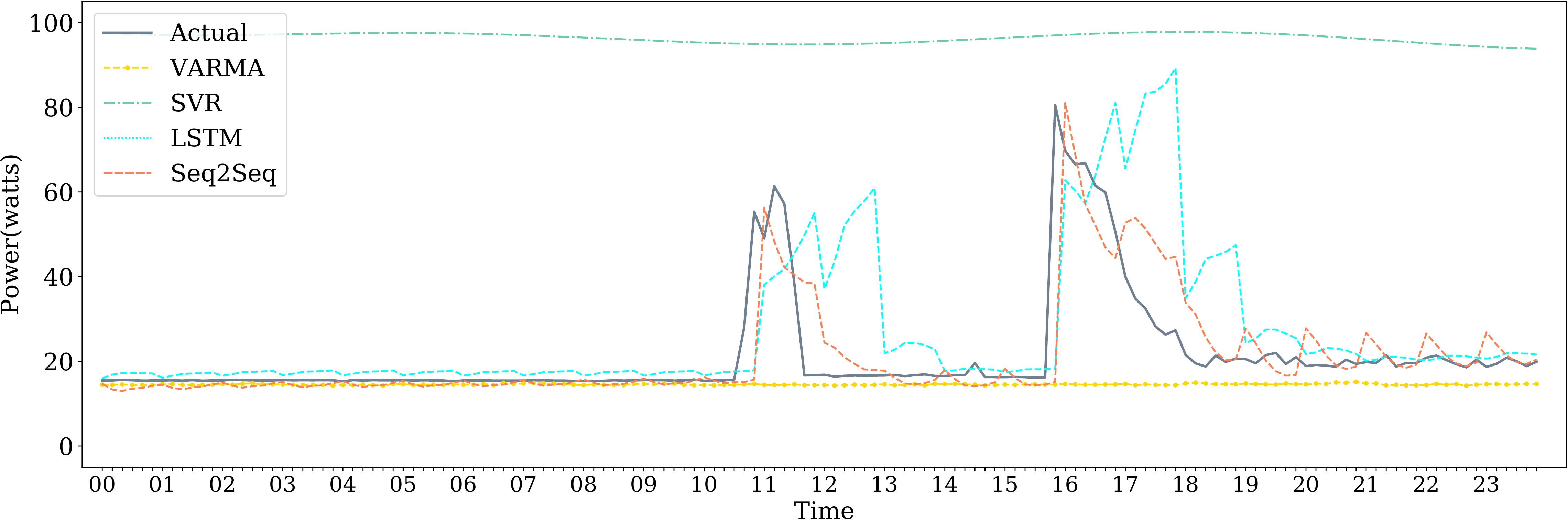}
        \subcaption{HiFi System and Routers}
        \label{fig:router}
    \end{subfigure}
    
    \begin{subfigure}{0.65\textwidth}
        \includegraphics[width=\linewidth]{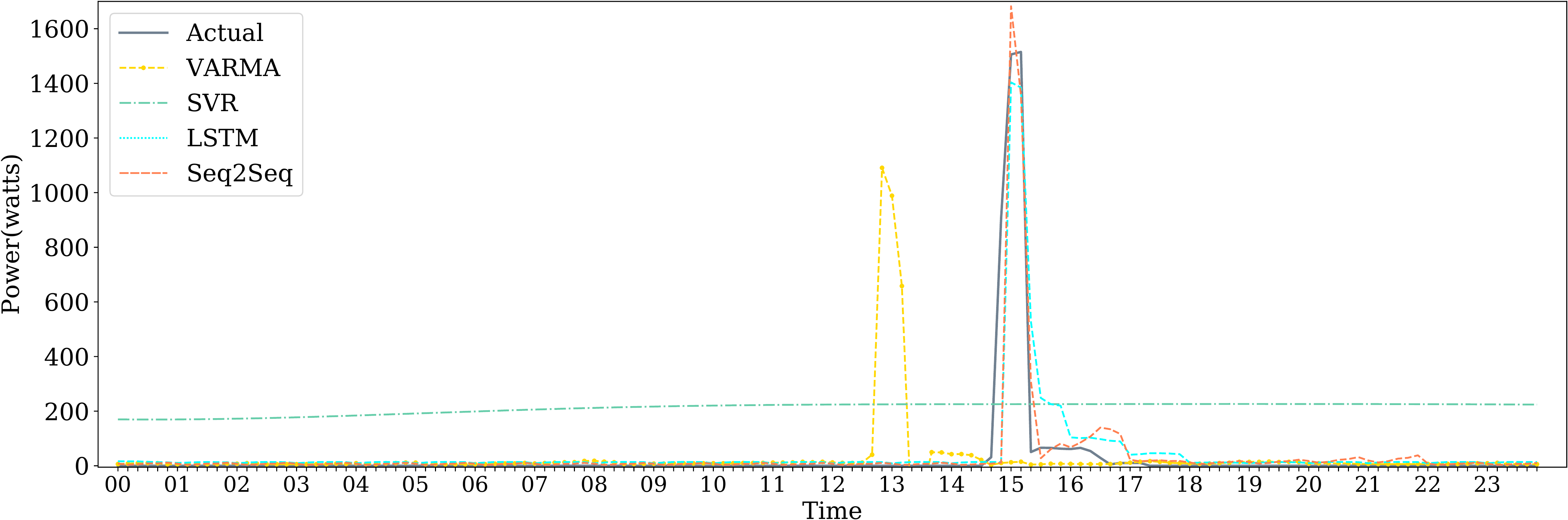}
        \subcaption{Dishwasher}
        \label{fig:dishwasher}
    \end{subfigure}
    
    \begin{subfigure}{0.65\textwidth}
        \includegraphics[width=\linewidth]{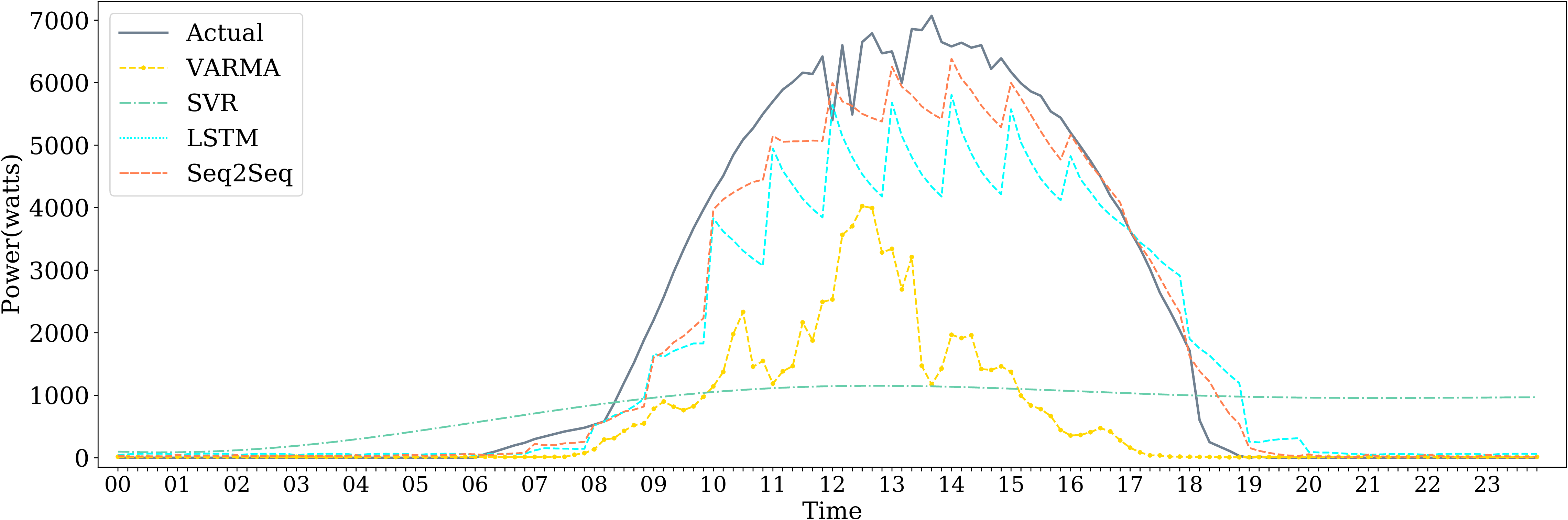}
        \subcaption{PV Power Production}
        \label{fig:pv}
    \end{subfigure}
    
    \begin{subfigure}{0.65\textwidth}
        \includegraphics[width=\linewidth]{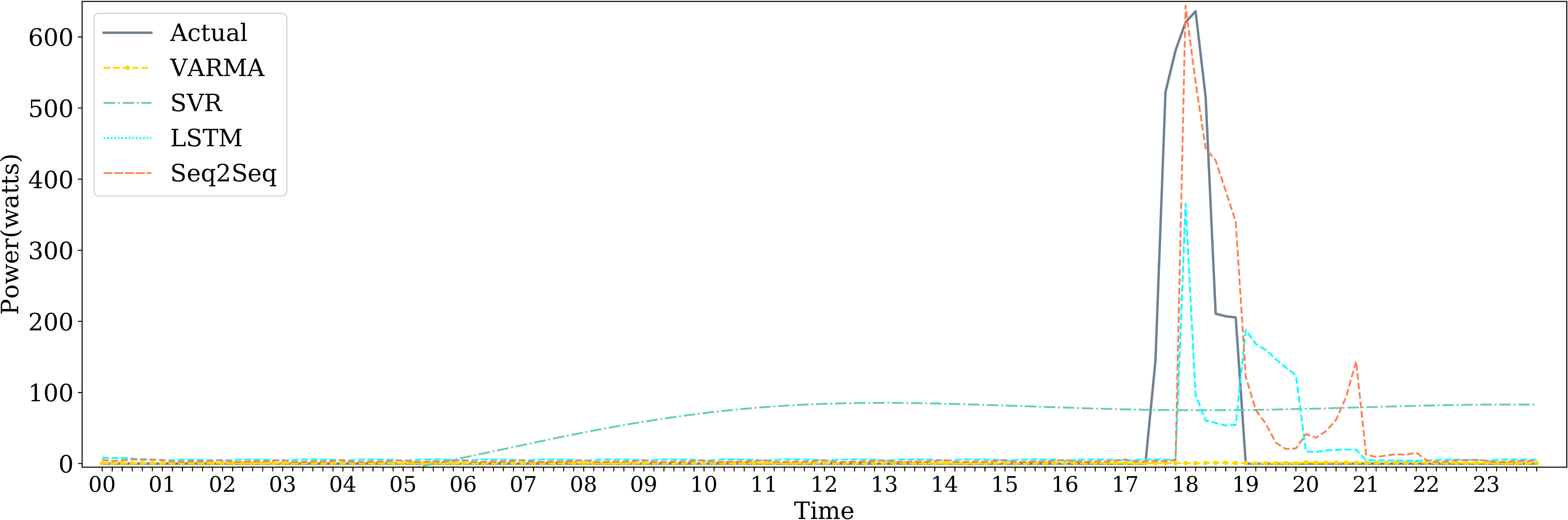}
        \subcaption{Tumble Dryer}
        \label{fig:dryer}
    \end{subfigure}
    
    \begin{subfigure}{0.65\textwidth}
        \includegraphics[width=\linewidth]{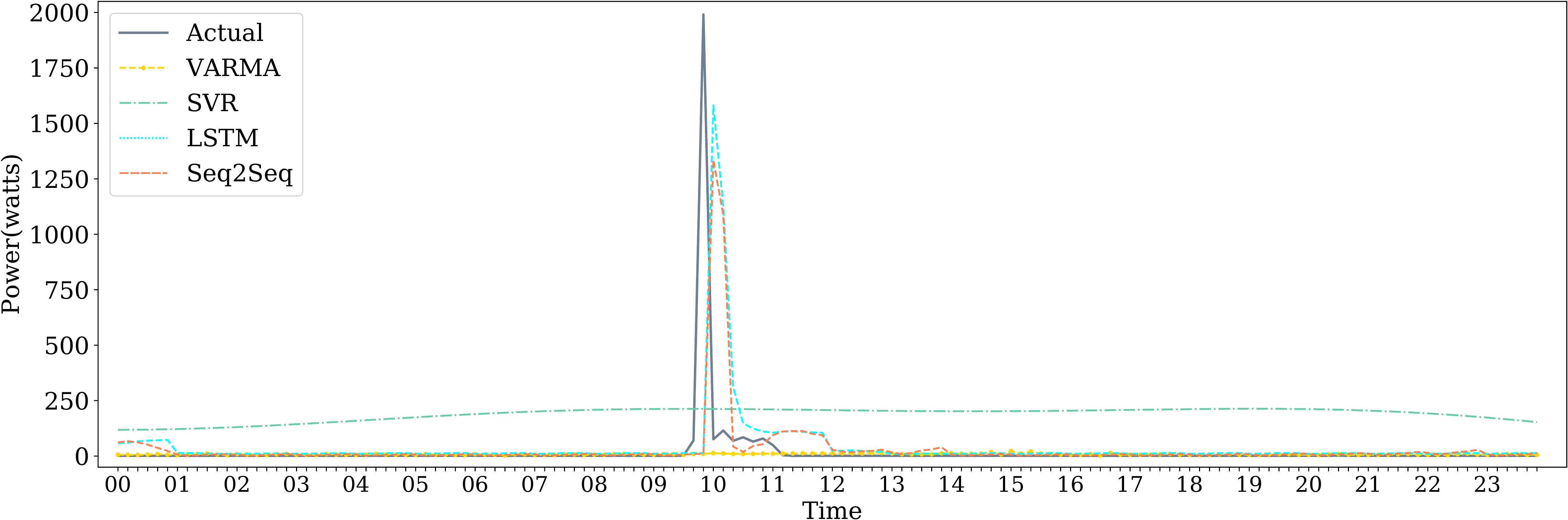}
        \subcaption{Washing Machine}
        \label{fig:washingmachine}
    \end{subfigure}
    \caption[]{Energy consumption prediction for 24 hours with a 10-minute resolution.}
    \label{fig:res}
\end{figure*}

RMSE, nRMSE, and wMAPE are calculated based on the prediction results of the baseline models and presented in Tables~\ref{tab:HouseDrmse}--\ref{tab:HouseDwmape}. Figures~\ref{fig:res}.\subref{fig:router}--\subref{fig:washingmachine} demonstrates load predictions versus actual load demand for HiFi system
and routers, dishwasher, PV power production, tumble dryer, washing machine, and overall energy consumption during a typical day using VARMA, SVR, 
LSTM and Seq2Seq models. Forecast results illustrate that the Seq2Seq model has the lowest error rate among all four models; in other words, it provides closer predictions to the actual demand. However, the RMSE and nRMSE of the LSTM model for tumble dryer and washing machine is 2.5\% and 0.5\% lower than the  Seq2Seq model, which is insignificant considering the fact that the Seq2Seq has lower wMAPE for both mentioned appliances.

\begin{table}[]
\centering
\vspace{1.5mm}
\caption{RMSE results of different model predictions for each appliance. The bold indicates the best performance.}
\label{tab:HouseDrmse}
\resizebox{\columnwidth}{!}{%
{\begin{tabular}{@{}lccccc@{}}
\toprule
\multicolumn{1}{c}{\textbf{Appliance}} & \textbf{VARMA} & \textbf{SVR} 
& \textbf{LSTM} & \textbf{Seq2Seq} \\ \midrule
HiFi System and Router  & 31.53  & 65.50  
& 12.38           & \textbf{11.58}  \\
Dishwasher             & 203.90 & 248.93 
& 162.24          & \textbf{159.99} \\
PV Power Production     & 1933.6 & 2086.1 
& 879.65          & \textbf{683.69} \\
Tumble Dryer            & 97.73  & 90.92  
& \textbf{53.50}  & 54.88           \\
Washing Machine         & 198.92 & 232.16 
& \textbf{149.41} & 150.25          \\
Total Power Consumption & 1588.1 & 1444.6 
& 1025.7          & \textbf{1017.2} \\ \bottomrule
\end{tabular}%
}}
\end{table}

\begin{table}[]
\centering
\caption{nRMSE results of different model predictions for each appliance. The bold indicates the best performance.}
\label{tab:HouseDnrmse}
\resizebox{\columnwidth}{!}{%
{\begin{tabular}{@{}lccccc@{}}
\toprule
\multicolumn{1}{c}{\textbf{Appliance}} & \textbf{VARMA} & \textbf{SVR} 
& \textbf{LSTM} & \textbf{Seq2Seq} \\ \midrule
HiFi System and Router  & 0.125 & 0.259 
& 0.049          & \textbf{0.045} \\
Dishwasher             & 0.106 & 0.129 
& 0.084          & \textbf{0.083} \\
PV Power Production     & 0.187 & 0.202 
& 0.085          & \textbf{0.066} \\
Tumble Dryer            & 0.117 & 0.109 
& \textbf{0.064} & 0.066          \\
Washing Machine         & 0.097 & 0.113 
& \textbf{0.072} & 0.073          \\
Total Power Consumption & 0.133 & 0.121 
& 0.085          & \textbf{0.085} \\ \bottomrule
\end{tabular}%
}
}
\end{table}

\begin{table}[]
\centering
\caption{wMAPE results of different model predictions for each appliance. The bold indicates the best performance.}
\label{tab:HouseDwmape}
\resizebox{\columnwidth}{!}{%
{\begin{tabular}{@{}lccccc@{}}
\toprule
\multicolumn{1}{c}{\textbf{Appliance}} & \textbf{VARMA} & \textbf{SVR} 
& \textbf{LSTM} & \textbf{Seq2Seq} \\ \midrule
HiFi System and Router  & 0.431 & 28007.08  
& 0.263 & \textbf{0.187} \\
Dishwasher             & 1.965 & 74463.87  
& 1.579 & \textbf{1.286} \\
PV Power Production     & 0.816 & 10238.64  
& 0.378 & \textbf{0.278} \\
Tumble Dryer            & 1.760 & 51988.42  
& 1.216 & \textbf{1.058} \\
Washing Machine         & 2.350 & 103436.34 
& 1.743 & \textbf{1.418} \\
Total Power Consumption & 0.952 & 10240.58  
& 0.546 & \textbf{0.492} \\ \bottomrule
\end{tabular}%
}}
\end{table}

\subsection{Operation Results}
\begin{figure}
\vspace{-10pt}
\centering
\includegraphics[width=8.5cm,height=4cm]{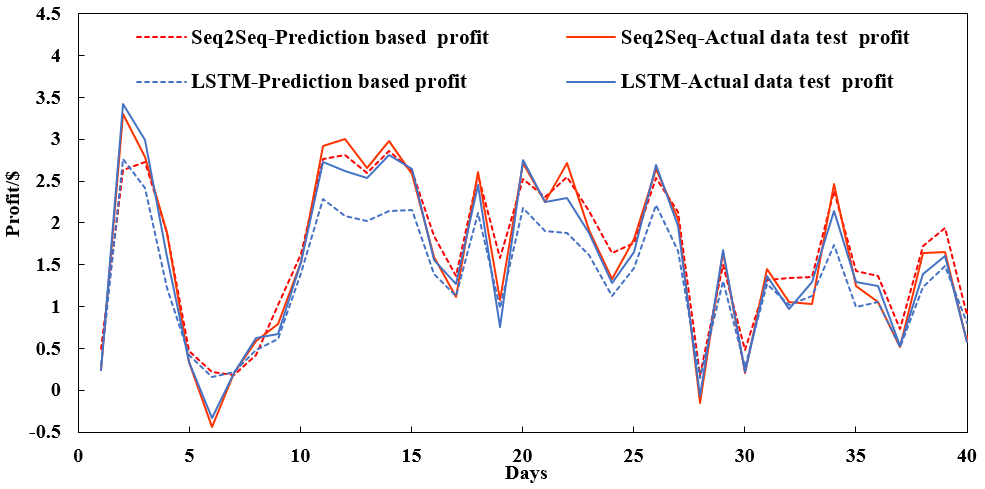}
\setlength{\abovecaptionskip}{0pt} 
\caption{Operation comparison of LSTM and Seq2Seq.}
\label{fig1}
\vspace{-10pt}
\end{figure}

In this section, we will show the HEMS operation results under different prediction methods. Based on the PV and load prediction results of each day, Algorithm~\ref{Q-LearningAlgo} produces both predicted and actual daily profit. This simulation is implemented in MATLAB, and each day is repeated ten runs to achieve the average value. Fig.~\ref{fig1} shows the 40 days prediction-based results and actual data test results of Seq2Seq and LSTM. The prediction-based results and actual data test results of Seq2Seq are close, while the prediction-based profit of LSTM is relatively lower than actual data test results.

\begin{figure}
\vspace{-5pt}
\centering
\includegraphics[width=8.5cm,height=4cm]{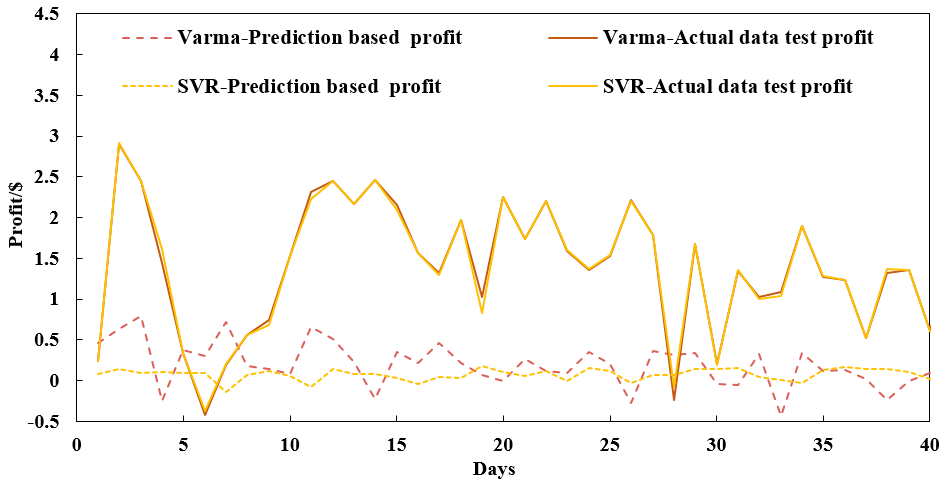}
\setlength{\abovecaptionskip}{0pt} 
\caption{Operation comparison of VARMA and SVR.}
\vspace{-10pt}
\label{fig2}
\end{figure}

\begin{figure}
\centering
\includegraphics[width=8.5cm,height=4cm]{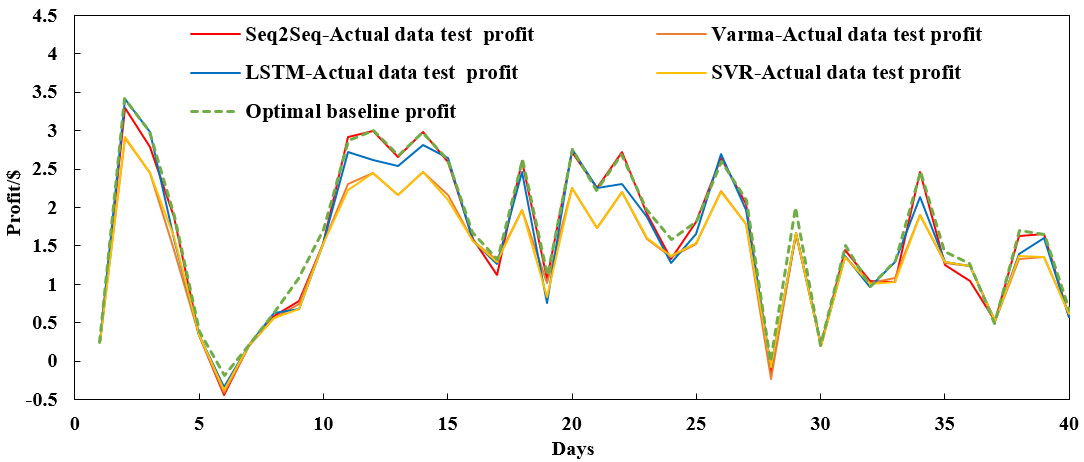}
\setlength{\abovecaptionskip}{0pt} 
\caption{40 days actual data test results comparison.}
\vspace{-5pt}
\label{fig3}
\end{figure}

Fig.~\ref{fig2} presents the performance of VARMA and SVR, in which both algorithms have a noticeable error between prediction-based profit and actual data test profit. This error can be explained by Fig.~\ref{fig:pv}, where the VARMA and SVR have a much lower predicted power for PV. The lower predicted power naturally leads to lower prediction-based results. However, in the actual data test period, the actual PV power is much higher, and a high profit is observed. Fig.~\ref{fig3} presents the actual data test results of different forecasting methods, which is the primary concern of consumers. We also used the actual data for Q-learning-based HEMS as an optimal baseline, which means a prediction method with no error. Seq2Seq and LSTM based HEMS benefit from the high forecasting accuracy, and a higher profit is observed. The Seq2Seq learning method has a comparable profit curve with an optimal baseline, while VARMA and SVR show a lower overall profit. On days 6, 28, 30, and 37, all algorithms have a comparable performance. A very low PV generation can explain those days, leading to a narrow profit difference in actual operation.

In summary, the Seq2Seq learning outperforms other algorithms with actual data test performance and a narrow error between prediction-based results. LSTM also results in good operation performance. VARMA and SVR have a significant error between prediction-based and actual data results.

\vspace{0.1in}

\section{Conclusion}
\label{Conclusion}
In this paper, we proposed a sequence-to-sequence learning-based HEMS. The proposed Seq2Seq model forecasts energy consumption for home appliances and the PV power production for the next hour with a 10-minute resolution. The forecasting accuracy of the proposed model is compared against VARMA, SVR, and LSTM on a real-world dataset. Evaluation results illustrate the superior performance of the proposed model. The outputs of the models are further fed into a Q-learning model for offline HEMS optimization, and the online performance of the optimization model is tested using the ground-truth values. Operation results indicate that the proposed model outperforms other algorithms, with its narrow error margin between online and offline operations.

\vspace{0.1in}

\section*{Acknowledgement}
Hao Zhou and Melike Erol-Kantarci were supported by the Natural Sciences and Engineering Research Council of Canada (NSERC), Collaborative Research and Training Experience Program (CREATE) under Grant 497981 and Canada Research Chairs Program.

\end{document}